\documentclass[aps,prl,twocolumn,superscriptaddress,tight]{revtex4-1}
\usepackage{amssymb}
\usepackage{amsfonts}
\usepackage{color}
\usepackage{graphicx}
\usepackage{amsmath}
\usepackage{times}
\usepackage{bm}
\usepackage{float}
\usepackage{lipsum}
\usepackage{import}
\usepackage{enumerate}
\usepackage[colorlinks,linkcolor=blue,anchorcolor=blue,citecolor=blue]{hyperref}
\makeatletter

\begin{document}

\title{Versatile Reservoir Computing for Heterogeneous Complex Networks}

\author{Yao Du}
\affiliation{School of Physics and Information Technology, Shaanxi Normal University, Xi'an 710062, China}

\author{Huawei Fan}
\affiliation{School of Science, Xi'an University of Posts and Telecommunications, Xi'an 710121, China}

\author{Xingang Wang}
\email{Corresponding author: wangxg@snnu.edu.cn}
\affiliation{School of Physics and Information Technology, Shaanxi Normal University, Xi'an 710062, China}

\begin{abstract}
A new machine learning scheme, termed versatile reservoir computing, is proposed for sustaining the dynamics of heterogeneous complex networks. We show that a {\it single, small-scale} reservoir computer trained on time series from a subset of elements is able to replicate the dynamics of {\it any} element in a large-scale complex network, though the elements are of different intrinsic parameters and connectivities. Furthermore, by substituting failed elements with the trained machine, we demonstrate that the collective dynamics of the network can be preserved accurately over a finite time horizon. The capability and effectiveness of the proposed scheme are validated on three representative network models: a homogeneous complex network of non-identical phase oscillators, a heterogeneous complex network of non-identical phase oscillators, and a heterogeneous complex network of non-identical chaotic oscillators.

\end{abstract}
\maketitle

{\it Background.--} Many real-world complex systems are composed of a large number of dynamical elements, with the system functionalities hinging on the collective behavior of the interacting elements~\cite{CN:RA,CN:Newman}. The sensitive nature of the dynamics of complex systems requires that for the systems to maintain their normal functions, all the elements should be working properly and coordinately~\cite{CN:JAA,REV:FAR}. For instance, the failure of a single station in the power grid, if not rectified promptly, might trigger a cascade of failures and finally result in a large-scale blackout~\cite{Cascade:LYC,Cascade:RA,Cascade:SVB}. This poses a significant challenge to the management of complex systems, as temporary and sporadic element failures are inevitable in real scenarios. The traditional approach to addressing this issue is preparing a hardware backup for each element and replacing the malfunctioning ones with backups in case of emergencies. With the advent of digital twins in machine learning~\cite{DT:TF,DT:LR,DT:BP,DT:KLW}, it is now possible to implement this approach virtually, in the sense that the failed elements can be substituted by artificial machines trained on measured data. 

From a dynamical perspective, if artificial machines can replicate the behavior of individual elements faithfully, the collective dynamics of the complex system can be maintained by replacing elements with machines, thereby preserving its overall functions. However, the success of this virtual approach relies on the accurate replication of the dynamics of all individual elements. For heterogeneous complex systems consisting of non-identical elements, this necessitates the construction and training of an ensemble of distinct machines, with each machine mimicking the dynamics of a specific element. As the machines are trained individually, such an approach involves a massive amount of data resources and demands an extremely high computational capacity, making it difficult to be implemented in large-scale complex systems. Moreover, for real-world complex systems, full observation of all elements is typically infeasible; often, what we can access is only time series acquired from a small subset of the elements, making it infeasible to learn the dynamics of the unobserved elements by machines directly. A question of both theoretical and practical significance thus is: Can a single, versatile machine be trained on time series of a few observable elements, such that the machine is able to substitute any element in a heterogeneous complex system while preserving its collective dynamics over a finite time horizon? Leveraging the techniques of reservoir computing (RC) in machine learning~\cite{RC:Maass2002,RC:Jaeger,RC:Pathak2017,RC:NC2024perspective,RC:adaptiveRCLYC}, we demonstrate that such a versatile machine can indeed be constructed, and it performs effectively in sustaining the collective dynamics of general heterogeneous complex systems.

\begin{figure*}[tbp]
\begin{center}
\includegraphics[width=0.95\linewidth]{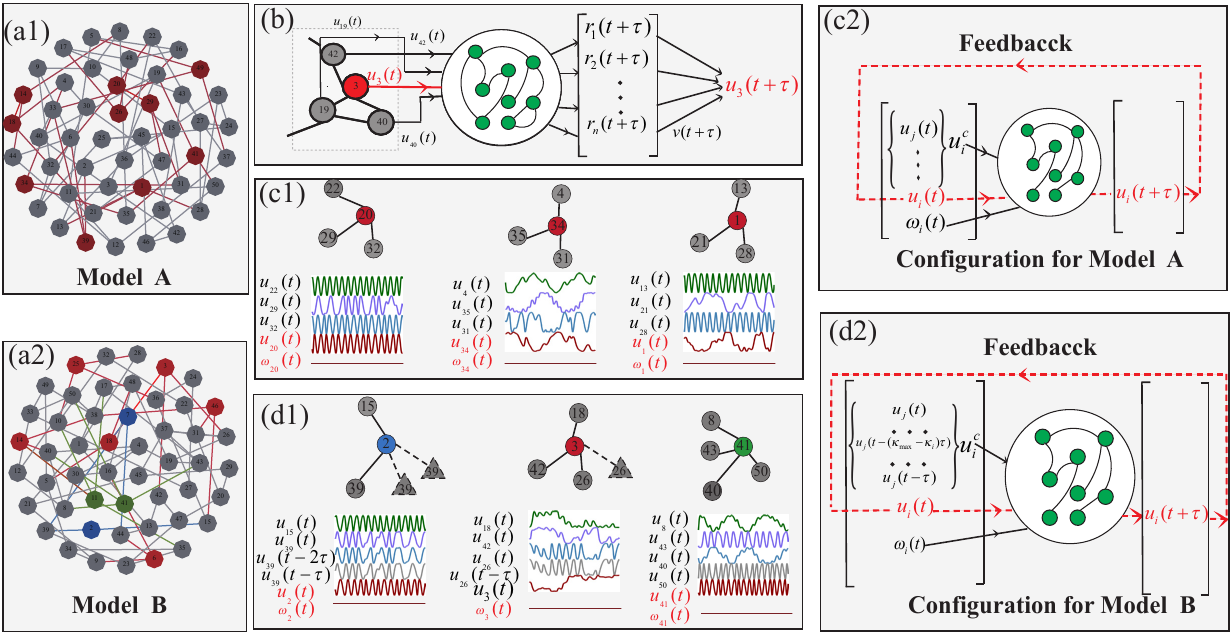}
\caption{Sustaining the dynamics of heterogeneous complex networks using versatile RC. (a1) Homogeneous complex network of $N=50$ non-identical phase oscillators (Model A), all with degree $k=3$. Red nodes ($m=10$) denote sampled oscillators used for machine training. (a2) Heterogeneous complex network of $N=50$ non-identical phase oscillators (Model B). The oscillators are of different degrees. Sampled oscillators ($m=10$) are colored by degrees: blue for $k=2$, red for $k=3$, and green for $k=4$. (b) Schematic of the scheme of parallelized RCs for homogeneous networks. (c1) Input data for Model A. (c2) Closed-loop configuration applied to Model A. (d1) Input data for Model B. Triangles represent the virtual oscillators created by the padding technique.  (d2) Closed-loop configuration applied to Model B.}
 \label{fig1}
\vspace{-0.5cm}
\end{center}
\end{figure*}

{\it Problem description.--} The dynamics of many complex systems can be modeled by networked oscillators as~\cite{CN:JAA,REV:FAR}
\begin{equation}
\dot{\mathbf{x}}_i=\mathbf{F}(\mathbf{x}_i,\beta_i)+\varepsilon\sum_{j=1}^N a_{ij} \mathbf{H}(\mathbf{x}_j,\mathbf{x}_i),
\end{equation}
where $i,j=1,\ldots,N$ are the oscillator indices, $\mathbf{x}_i$ is the state vector of oscillator $i$, $\mathbf{F}(\mathbf{x}_i,\beta_i)$ denotes the dynamics of oscillator $i$ in the isolated form, $\mathbf{H}(\mathbf{x}_j,\mathbf{x}_i)$ describes the coupling function, and $\varepsilon$ represents the uniform coupling strength. The coupling relationship of the oscillators is encoded by the adjacency matrix $\mathbf{A}$: $a_{ij}=a_{ji}=1$ if oscillators $i$ and $j$ are connected, and $a_{ij}=0$ otherwise. The number of connections (degree) associated with oscillator $i$ is $k_i=\sum_j a_{ij}$. Heterogeneity may arise from either the oscillator dynamics (where the intrinsic parameter $\beta_i$ differs among oscillators) or the local connectivity (where oscillators have varying degrees). We consider the scenario that both the network structure and the oscillator intrinsic parameters are known {\it a priori}, but the equations governing the system dynamics are unknown. We assume further that time series can be measured from only a small subset of the oscillators ($m$ in total, with $m\ll N$). Our primary objective is to design and train a versatile machine that can replace any oscillator while maintaining the system dynamics for an extended period.

Three representative network models are adopted in our studies to validate the capability and effectiveness of the proposed machine learning scheme: a homogeneous complex network of non-identical phase oscillators [Model A, Fig.~\ref{fig1}(a1)], a heterogeneous complex network of non-identical phase oscillators [Model B, Fig.~\ref{fig1}(a2)], and a small-size complex network of non-identical chaotic oscillators (Model C). Weak couplings are employed in all network models to ensure that the oscillators are weakly correlated and the systems exhibit spatiotemporal chaotic behaviors [see Supplementary Materials (SM) for details on the network models].

{\it Method.--} The architecture of versatile RC is the same as that of standard RC~\cite{RC:Maass2002,RC:Jaeger,RC:Pathak2017,RC:NC2024perspective,RC:adaptiveRCLYC}, except for an expanded input layer. More specifically, the new design integrates the schemes of parallelized RCs~\cite{RC:Pathak2018,RC:Parlitz2018,RC:AtmosphereForecastOtt2020,RC:ParalNGRC,RC:ParallelPRL2022} (proposed for learning spatially extended systems) and parameter-aware RC~\cite{KLW:2021,RC:Kim2021,RC:FHW2021,RC:XR2021,RC:ZH2021,RC:LHB2024} (developed for inferring unseen dynamics beyond the training set) by introducing two additional channels into the input layer, one for the coupling signals and the other for the control parameter. According to the network heterogeneities, the machine can be implemented in different configurations. Displayed in Fig.~\ref{fig1}(b) is the simplest configuration applicable to homogeneous networks of identical oscillators (the conventional scheme of parallelized RCs), where the input layer contains only the state and coupling channels. For this configuration, the input vector reads $\mathbf{U} = (\mathbf{u}_i, \mathbf{u}^c_i)$, with $\mathbf{u}_i$ being the state vector of the target oscillator (which is inputted into the reservoir through the state channel) and $\mathbf{u}^c_i = \{\mathbf{u}_j\}$ being the coupling signals that the target oscillator receives from its neighbors (which is inputted into the reservoir through the coupling channel). The training and validation procedures are the same as those of standard RC. In applications, we choose at random an oscillator in the network and substitute it with the machine. In substituting an oscillator, the state vector in the input channel is replaced by the machine output, while the coupling signals are still provided by the neighbors. Owing to the homogenous oscillator dynamics and connectivity, the target oscillator generating the training data can be selected at random, while the trained machine is capable of inferring the dynamics of any oscillator, provided the coupling signals are available. This configuration, however, applies to only homogeneous networks of identical degree and oscillator dynamics~\cite{RC:Pathak2018,RC:Parlitz2018,RC:AtmosphereForecastOtt2020,RC:ParalNGRC}, which do not align with the research interest of our current study.

Plotted in Fig.~\ref{fig1}(c2) is the configuration applicable to homogeneous complex networks of non-identical phase oscillators. In addition to the state and coupling channels, a parameter-control channel is introduced in the input layer in this configuration. Following the scheme of parameter-aware RC~\cite{KLW:2021,RC:Kim2021,RC:FHW2021,RC:XR2021,RC:ZH2021,RC:LHB2024}, we choose the natural frequencies, $\omega_i$, of the target oscillator as the input of the parameter-control channel. Different from the configuration shown in Fig.~\ref{fig1}(b), here the training data is a concatenation of the time series from several oscillators, as depicted in Fig.~\ref{fig1}(c1). The input vector has the form $\mathbf{U} = (\mathbf{u}_i, \mathbf{u}^c_i, \omega_i)$, with $\omega_i(t)$ being a step-function of time. The training and validation processes for this configuration are identical to those of parameter-aware RC. In applications, the machine receives not only the coupling signals from neighboring oscillators, but also the constant parameter $\omega_i$ associated with the target oscillator (see SM for details). Due to the heterogeneous oscillator dynamics, the performance of the machine is dependent on the sampled oscillators generating the training data (see SM for details).    

The configuration applicable to complex networks with both heterogeneous structure and oscillator dynamics is depicted in Fig.~\ref{fig1}(d2). For this type of complex network, the key issue is how to cope with the diverse degrees of the oscillators, so that the same machine can be used for substituting any oscillator. This issue can be addressed by introducing virtual oscillators as neighbors through a padding technique, as follows. Let $k_{max}$ be the largest degree of the oscillators and $k_i$ be the degree of the sampled oscillator $i$, we generate $L=k_{max}-k_i$ virtual oscillators by delaying the coupling signals from one of its neighbors. Illustrated in Fig.~\ref{fig1}(d1) is the schematic of the padding technique for typical oscillators with degrees smaller than $k_{max}$. Taking oscillator $2$ as an example, the oscillator is connected to oscillators $15$ and $39$ in the network and has the degree $k=2$. As the largest degree is $k_{max}=4$, we thus introduce two virtual oscillators as its new neighbors, one with the time series $\mathbf{u}_{39}(t-\tau)$ and the other with the time series $\mathbf{u}_{39}(t-2\tau)$, with $\tau$ being the time-delay coefficient. Virtual oscillators provide no new information about the system dynamics, but make the machine versatile, i.e., the dimension of the input vector $\mathbf{U}$ becomes identical across the oscillators. Treating virtual oscillators as real ones, the implementation of the machine under this configuration is identical to that of Model A (see SM for details). 

{\it Results.--} We present first the results for Model A depicted in Fig.~\ref{fig1}(a1). The coupling strength is chosen as $\varepsilon=0.4$, by which the largest Lyapunov exponent of the network is about $\Lambda=0.1$ (the corresponding Lyapunov time is about $10$) and the time-averaged synchronization order parameter is about $\left<R\right>=0.23$ (see SM for definition of $R$). Training data are acquired from $m=10$ oscillators chosen at random, and the time series acquired from each oscillator contains $5000$ data points. Each time series is divided into two segments. The first segment, containing $4000$ data points, is used for machine training, and the second segment, containing $1000$ data points, is used for machine optimization. The state of oscillator $i$ is represented by the vector $\mathbf{u}_i=[\sin(\theta_i),\cos(\theta_i)]$, which is also the coupling vector that oscillator $i$ sends to its neighbors. As the oscillators are of identical degree ($k=3$), the dimension of the input data is $D_{in}=2+2k+1=9$. The training and validation processes are identical to those of parameter-aware RC (see SM for details). 

\begin{figure}[tbp]
\begin{center}
\includegraphics[width=0.95\linewidth]{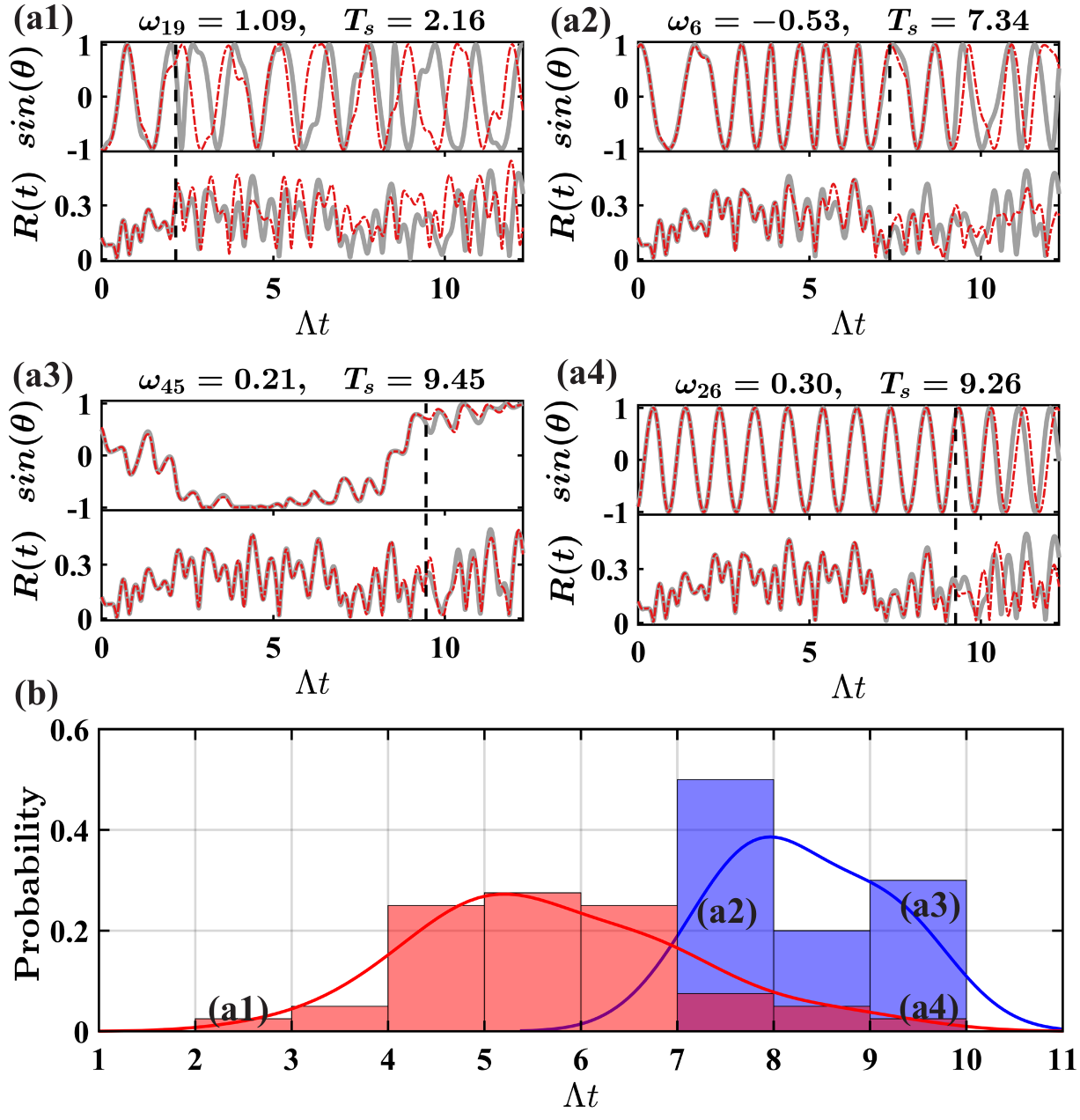}
 \caption{Performance of the machine in substituting a single oscillator in Model A. Shown in (a) are typical results for (a1) non-sampled oscillators with irregular motions, (a2) sampled oscillators with irregular motions, (a3) sampled oscillators with regular motions, and (a4) non-sampled oscillators with regular motions. Top panels: true (grey) vs. predicted (red) state evolutions. Bottom panels: time evolution of the synchronization order parameter, $R(t)$, for the original (grey) and substituted (red) networks. Dashed lines indicate maintenance horizons ($T_s$). (b) Distribution of maintenance horizons across all oscillators. Results for the sampled and non-sampled oscillators are colored in blue and red, respectively.}
 \vspace{-0.7cm}
\label{fig2}
\end{center}
\end{figure}

Shown in Fig.~\ref{fig2} is the performance of the machine in substituting a single oscillator in Model A. The machine performance is evaluated by two metrics: the accuracy in replicating the local dynamics of the substituted oscillator (the state evolution) and the capability in maintaining the collective behavior of the whole network [the synchronization order parameter $R(t)$]. In substituting an oscillator, we operate the machine first in the open-loop mode by inputting a short sequence acquired from the original system (to remove the impacts of the initial conditions of the reservoir), and then in the closed-loop mode as depicted in Fig.~\ref{fig1}(c2) (see SM for details). We see in Fig.~\ref{fig2} that the machine performance is of significant variation across the oscillators. Generally, the time horizon (see SM for definition) for which the network dynamics can be accurately maintained depends on two factors: whether the substituted oscillator is within the sampling set and whether it exhibits regular motion. Specifically, a better performance is achieved when the oscillator is within the sampling set and exhibitting regular motion [e.g. Fig.~\ref{fig2}(a3)], while the performance is worse if the oscillator is not included in the sampling set or presenting irregular motion [e.g. Fig.~\ref{fig2}(a1)]. The oscillator-dependent performance is understandable, as the machine is trained on the sampled oscillators, and replicating regular motions is relatively easier compared to irregular ones. On average, the maintenance horizon is about $6.2$ Lyapunov times, signifying the capability of the machine in maintaining the network dynamics when a single oscillator is substituted. 

Additional studies have also been conducted to check the dependence of the machine performance on other factors, including the size of the sampling set, the number of substitutions, and the uniform coupling strength. The general finding is that the machine performance is improved by adopting more sampled oscillators or by increasing the uniform coupling strength, and is declined if more oscillators are substituted. (Details are given in SM).

\begin{figure}[tbp]
\begin{center}
\includegraphics[width=0.95\linewidth]{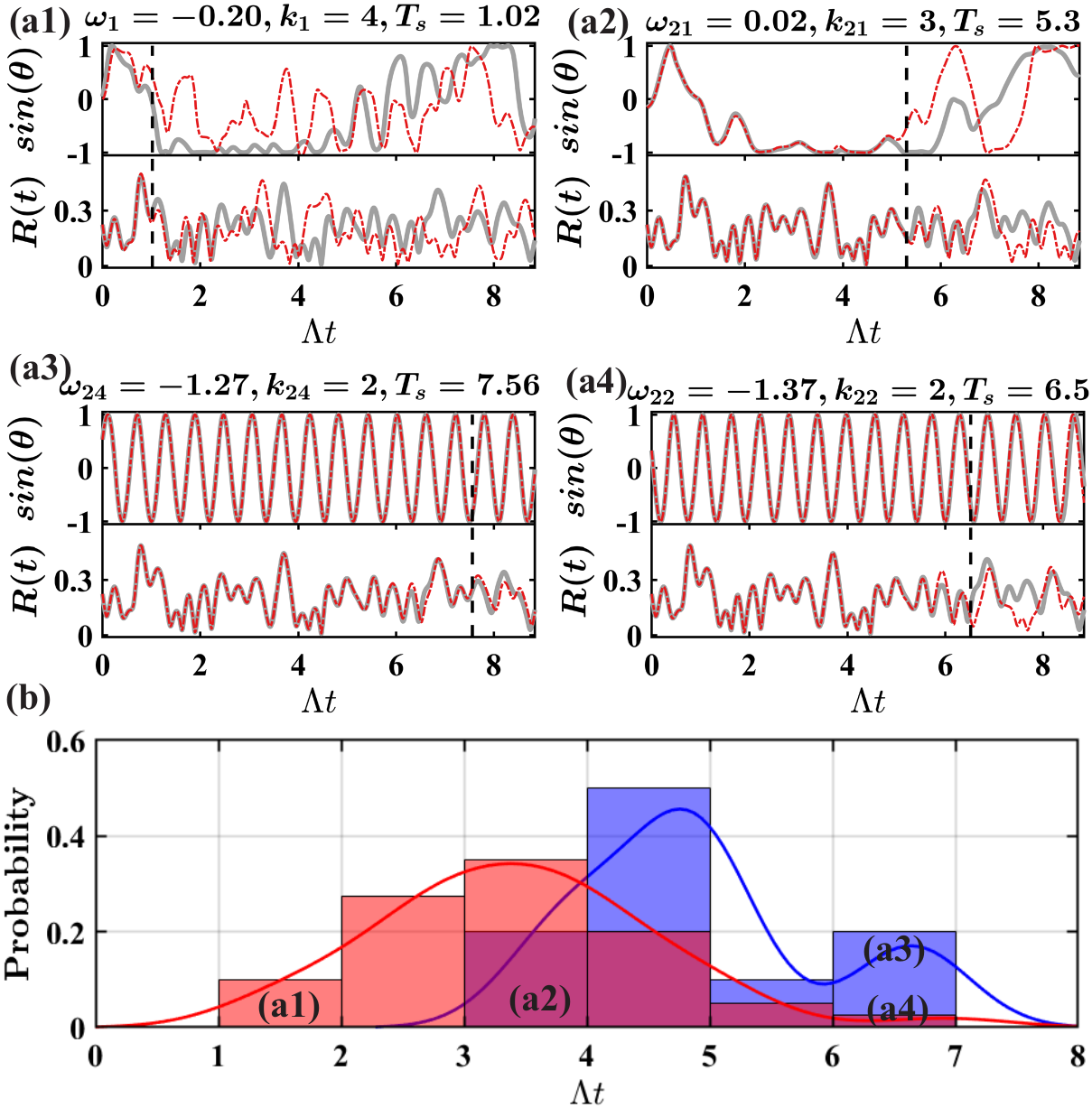}
 \caption{Performance of the machine in sustaining the dynamics of Model B. (a1–a4) Same as in Fig.~\ref{fig2} but for the heterogeneous network. Grey: original system; red: with substitution. (b) Distribution of maintenance horizons for sampled (blue) and non-sampled (red) oscillators.}
 \vspace{-0.7cm}
\label{fig3}
\end{center}
\end{figure}

We next present the results for Model B displayed in Fig.~\ref{fig1}(a2). The coupling strength is still chosen as $\varepsilon=0.4$, by which the largest Lyapunov exponent of the network is about $\Lambda=0.09$ (the corresponding Lyapunov time is about $11$) and the time-averaged synchronization order parameter is about $\left<R\right>=0.27$. Training data are also acquired from $m=10$ oscillators, with the degree of the sampled oscillators ranging from $2$ to $4$. By the padding technique depicted in Fig.~\ref{fig1}(d1), we introduce virtual oscillators to sampled oscillators with degrees smaller than $4$. As $k_{max}=4$, the uniform dimension of the input vector is $D_{in}=2+2k_{max}+1=11$. The training and validation processes are identical to those in Model A (see SM for details). The performance of the machine in substituting a single oscillator is shown in Fig.~\ref{fig3}. Similar to the findings in Model A, we see that better performance is achieved when the oscillator is within the sampling set and presenting regular motion [e.g. Fig.~\ref{fig3}(a3)], while worse performance is observed when the oscillator is not included in the sampling set or presenting irregular motion [e.g. Fig.~\ref{fig3}(a1)]. For Model B, the average maintenance horizon is approximately $5.1$ Lyapunov times, which is slightly lower than that of Model A.

\begin{figure}[tbp]
\begin{center}
\includegraphics[width=0.98\linewidth]{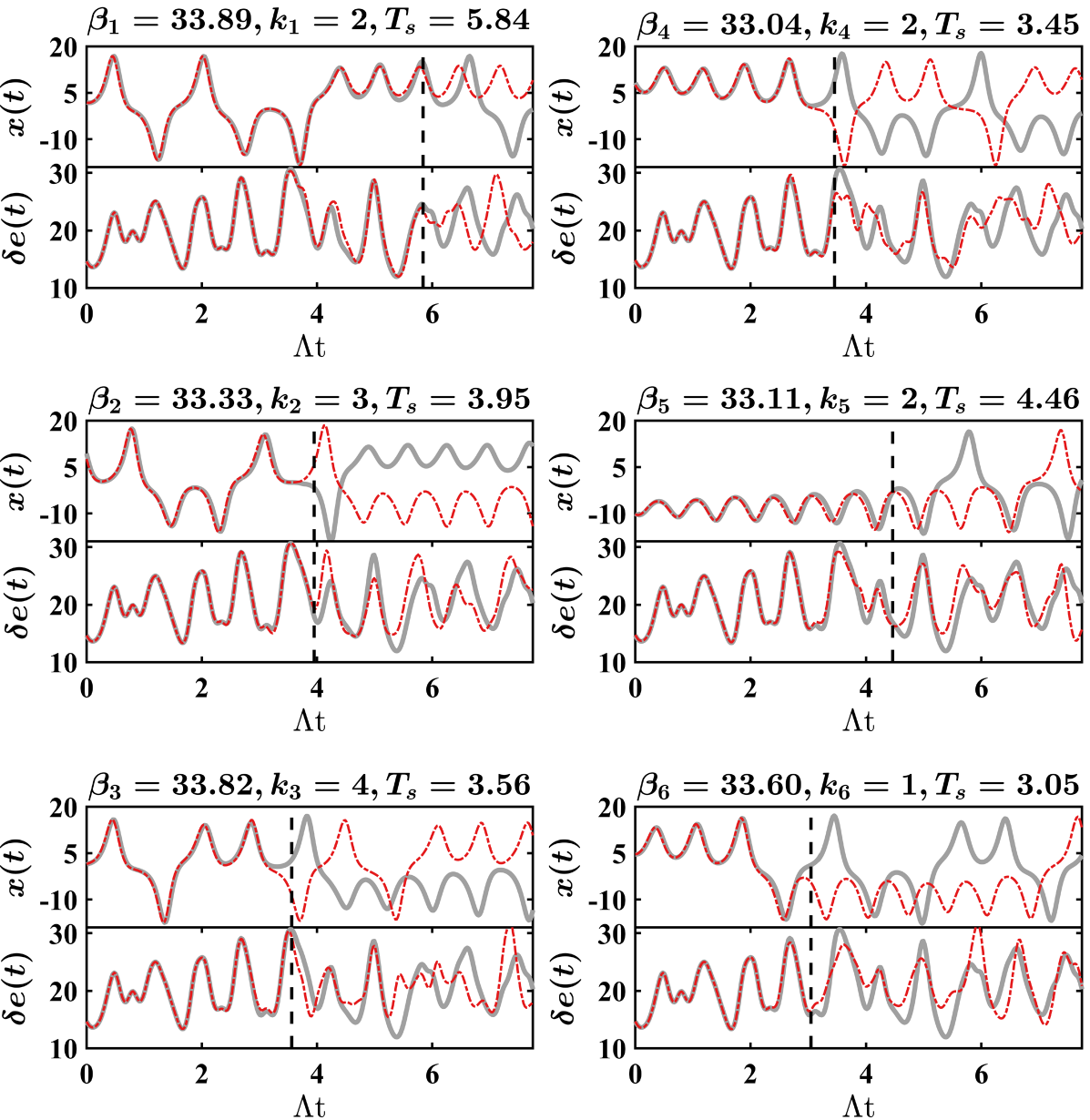}
 \caption{Performance of the machine in maintaining the dynamics of Model C. Results for the sampled and non-sampled oscillators are shown in the left and right columns, respectively. In each subplot, the top and bottom panels display, respectively, the state evolution of the substituted oscillator and the time evolution of the network synchronization error, $\delta e$. Grey curves are results for the original network. Red curves are results with substitution. Vertical dashed lines denote the maintenance horizons.}
 \vspace{-0.6cm}
\label{fig4}
\end{center}
\end{figure}

We finally present the results for a small complex network of non-identical chaotic oscillators (Model C). The network is constructed by $6$ chaotic Lorenz oscillators and $7$ links. The oscillators are of different intrinsic parameters and degrees (ranging from $1$ to $4$), and are coupled through their $x$ variables in a linear fashion (see SM for details). A weak coupling strength is chosen to ensure the network is fully desynchronized and presents spatiotemporal chaotic behaviors. The largest Lyapunov exponent of the network is about $1.1$; the corresponding Lyapunov time is about $0.9$. Training data are acquired from $m=3$ oscillators (one with degree $k=2$, one with degree $k=3$, and one with degree $k=4$). The state vector of oscillator $i$ is $\mathbf{u}_i=[x_i,y_i,z_i]$, and the coupling signal that oscillator $i$ receives from oscillator $j$ is $u_j=x_j-x_i$. Again, padding technique is employed to expand the dimension of the input vector. As $k_{max}=4$, in this case the uniform dimension of the input vector is $D_{in}=3+4+1=8$ (the additional dimension accounts for the intrinsic parameter inputted into the reservoir through the parameter-control channel). The training and validation processes are the same as those of Models A and B (see SM for details). Shown in Fig.~\ref{fig4} is the performance of the machine in maintaining the network dynamics when a single oscillator is substituted. We see that both the dynamics of the substituted oscillator and the collective dynamics of the whole network [characterized by the synchronization error $\delta e(t)= \sum_{i,j=1}^N\left\| \mathbf{u}_i(t) - \mathbf{u}_j(t) \right\|/N^2$] are accurately sustained for a certain period (ranging from $3$ to $6$ Lyapunov times).  

{\it Discussions.--} While the power of RC in inferring chaotic dynamics has been well established, its application to complex systems with heterogeneous dynamics and structures remains a significant challenge. Unlike existing approaches that seek to replicate the dynamics of all individual elements in a complex system~\cite{RC:Pathak2018,RC:Parlitz2018,RC:AtmosphereForecastOtt2020,RC:ParalNGRC,RC:ParallelPRL2022}, our present work focuses on preserving the collective behavior of a complex network by substituting the failed element with a versatile machine. The new machine is distinguished by three key features: (1) it is compact; (2) it is trained on partial observations; and (3) it can substitute any element, though the elements are of different intrinsic parameters and connectivity. The success of the new machine hinges on its ability to transfer knowledge from the sampled oscillators to non-sampled ones, which is accomplished by integrating a parameter-control channel (to accommodate differences in oscillator intrinsic parameters) and a coupling channel (to account for variations in oscillator degree) into the standard RC architecture. The all-in-one and plug-and-play nature of versatile RC offers substantial convenience for maintaining the functionality of real-world complex systems suffering sporadic or temporary component failures. The convenience, however, comes at the cost of {\it a priori} knowledge about the system dynamics, including the intrinsic oscillator parameters, the local connectivities, and a short ``warm-up" time series for initializing the reservoir. The first two requirements might be met by the techniques of parameter estimation and structure identification developed in complex systems~\cite{Parameter:Parlitz,Network:DZR,Network:LYC}, while the third might be mitigated by adopting alternative RC architectures such as the next-generation RC~\cite{RC:DJG2021}.

The data and codes used in the current study can be found in Ref.~\cite{Github}.

\begin{acknowledgments}
This work was supported by the National Natural Science Foundation of China under Grant No.~12275165.
\end{acknowledgments}

\end{document}